\documentclass
[twocolumn,pra,superscriptaddress,showpacs]{revtex4-1}
\usepackage{amsmath,multirow,graphicx,amssymb,dcolumn,mathrsfs,bm,color,leftidx,array}
\makeatletter

\newcommand{\Rmnum}[1]{\expandafter\@slowromancap\romannumeral #1@}
\makeatother

\begin{document}

\title{Unconventional steady states and topological phases in an open two-level non-Hermitian system}
\author{Jian Xu}
\email{xujian$_$328@163.com;[Jian Xu and Youneng Guo contributed equally to this work]}\affiliation{College of Electronics and Information Engineering, Guangdong Ocean University, Zhanjiang, 524088,
China}

\author{Youneng Guo}
\email{guoxuyan2007@163.com;[Jian Xu and Youneng Guo contributed equally to this work]}\affiliation{Interdisciplinary Center for Quantum Information, National University of Defense Technology, Changsha 410073, People's Republic of China}\affiliation{School of Electronic Information and Electrical Engineering, Changsha University, Changsha, Hunan
410022, People's Republic of China}

\begin{abstract}

Decoherence and non-Hermiticity are two different effects of the open quantum systems. Both of them have
triggered many interesting phenomena. In this paper, we theoretically study an open two-level non-Hermitian system coupling to a dissipative environment by solving the vectorized Lindblad equation. This scheme provides us a powerful framework to address widespread open systems with gain, loss and dissipation. Our results show that there exist a new class of exceptional points (EPs) and steady states due to the interplay between non-Hermiticity and decoherence. Furthermore, we also demonstrate a new-type topological properties of eigenstates with zero real-part of eigenvalues ($Re[\lambda]=0$) which are corresponding to Fermi arcs. It is revealed that the phases of eigenstates located in Fermi arcs regime have a topological phase $|\pi/2|$ which is totally unaffected by the dissipative environment. Our results provide a promising approach for further uncovering and understanding the intriguing properties of non-Hermitian open systems.

\end{abstract}


\maketitle

\section{Introduction}

Decoherence and non-Hermiticity are two different effects to describe the open quantum systems giving rise to a great variety of exotic physical phenomena. The effect of decoherence caused by unavoidable coupling with the environment, resulting in the loss of energy, coherence, and information into the environment is ubiquitous in open quantum systems~\cite{Breuer}. It is widely accepted that the decoherence is a major obstacle to implement quantum computing and quantum information processing, which rely on keeping the evolution of quantum coherent undisturbed by the environment. Due to the fact that the decoherence effect is viewed as an undesirable destructing factor, tackling decoherence is a critical issue faced in the area of quantum engineering. Massive efforts have been payed to overcome decoherence by using entanglement distillation protocol~\cite{Kwiat}, decoherence-free subspace~\cite{Lidar}, quantum Zeno effect~\cite{Facchi}, and so on.

Compared to the decoherence effect, the non-Hermiticity refers to the system of interest with non-Hermitian Hamiltonian, which is a physical paradigm for the
description of dissipative quantum systems. One of the interesting phenomena induced by non-Hermiticity is the emergence of exceptional points (EPs)~\cite{Heiss,Doppler}, which are referred to not only the degeneracies of energy spectra but also the coalescence of eigenfunctions. In recent years, the concept of EPs has attracted extensive research interest in theory and experiment~\cite{Dembowski,Chen,Zhang,Dora,Cerjan,Kawabata,Yoshida,Weidemann}. As a result, many new features inaccessible in Hermitian systems can be observed in non-Hermitian systems, e.g., the violation of no-signaling principle~\cite{Lin,Regensburger}, loss-induced lasing~\cite{Brandstetter,Peng}, and the optimal brachistochrone problem~\cite{Huang}, and so on.

Beyond EPs, the properties of the spectra of non-Hermitian systems have also attracted much interest in the last decades~\cite{Hodaei,Xiao,Chen2,Ozdemir,Ghatak,Chu}. In particular, the notion of parity-time (PT) symmetry characterizing a wide class of non-Hermitian Hamiltonians have purely real spectra.~\cite{AGuo}. On the other hand, the non-Hermitian Hamiltonians with zero real (imaginary) parts of spectra are corresponding to Fermi (i-Fermi) arcs~\cite{Bergholtz} which can result in an emergent Riemann sheet topology~\cite{Berry}. Indeed, Fermi (i-Fermi) arcs are not only associated with the system state jumping from one eigenstate to another~\cite{Heiss2,Wang} but also deeply related to the source of novel bulk state~\cite{Kozii,Zhou}, new gapless phases~\cite{Bessho}, knotted non-Hermitian nodal structures~\cite{Carlstrom} and so on.

It is clear that a time-independent Hamiltonian with null eigenvalue (not only the real part but also imaginary part of eigenvalue is zero) is related to the occurrence of steady state and play essential roles in quantum control and quantum phase transition~\cite{Johan,Pei}. A current topic is the design of a system Hamiltonian to obtain a desired steady state or steady-state property. Recently, a sufficient criterion for determining the steady state in dissipative Hermitian systems has been proposed~\cite{Mostafavi}. Whether this criterion is well-defined in non-Hermitian systems remains questionable due to the non-Hermiticity of Hamiltonians whose spectra are in general complex. Hence, in this particular case, the condition of steady state as well as steady-state property in open non-Hermitian systems deserve a dedicated study.

Taking the decoherence and non-Hermiticity effects into consideration together, novel features of open non-Hermitian systems are expected to occur.
In this paper, we address a general non-Hermitian open systems through the vectorized Lindblad equation, whose eigenvalues and eigenstates are analytically solvable. The vectorization frame of the Lindblad equation suggests that one can cast the Lindblad equation to a vectorized equation~\cite{Caspel,Arkhipov,Minganti3,Havel,Am-Shallem,Scopa,Hatano}. As an example, we apply this technique to an open two-level non-Hermitian system with dissipation and propose a general condition of steady states in open non-Hermitian systems. Our results show that there exist a new type of EPs and steady states due to the interplay between non-Hermiticity and decoherence. Moreover, we also study the topological properties of Fermi arcs and EPs. The result shows that the topological phases of EPs as a specialized case of ones in Fermi arcs are robust against the environmental decoherence, and possess a $|\pi/2|$ phase regardless of the dissipative parameter.

The contents of this paper are organized as follows. In section II, we firstly introduce the non-Hermitian vectorized Lindblad equation and generalize the condition of steady states to a general open non-Hermitian system. In section III, we consider an open two-level system with non-Hermiticity and dissipations and theoretically solve the vectorized Lindblad equation of the model system. The properties of steady states and the topological phases in the presented system are discussed. At last, we give a remark and conclusion in section IV.

\section{The non-Hermitian Lindblad equation and the condition of steady states for a general open system}

In this section, we extend the Lindblad master equation of a Hermitian open system to a non-Hermitian case by transforming density matrices into vectors. To do this, we are here interested in a general non-Hermitian open system with gain, loss and dissipation. In the limit of weak Markovian time-independent interaction, the time evolution of an open quantum system can be expressed by a non-Hermitian Lindblad-type master equation $(\hbar= 1)$~\cite{Walls}
\begin{equation}
\label{master equation}
\frac{\partial\rho(t)}{\partial t}=-i[\hat{H}\rho(t)-\rho(t)\hat{H}^\dag]+\sum_{\mu}D[\hat{\Gamma}_{\mu}]\rho(t).
\end{equation}
$\hat{H}$ is a non-Hermitian Hamiltonian ($\hat{H}\neq\hat{H}^{\dagger}$) of the system. $D[\Gamma_{\mu}]\rho(t)$ are the dissipators associated with the Lindblad operators $\hat{\Gamma}_{\mu}$ describing directly the dissipative effects of environment. Without loss of generality, the most general quantum dynamical semigroup form of the dissipator can be written as
\begin{equation}
\label{Lindblad operators}
D[\hat{\Gamma}_{\mu}]\rho(t)=\hat{\Gamma}_{\mu}\rho(t)\hat{\Gamma}_{\mu}^{\dag}-\frac{1}{2}\hat{\Gamma}_{\mu}\hat{\Gamma}_{\mu}^{\dag}\rho(t)-\frac{1}{2}\rho(t)\hat{\Gamma}_{\mu}\hat{\Gamma}_{\mu}^{\dag}.
\end{equation}
The first term in Eq.(2) is called a quantum jump term, and the remaining terms are called continuous non-unitary dissipation terms~\cite{Ju}, respectively. Note that the non-Hermitian Hamiltonian $\hat{H}$ is quite different from an effective non-Hermitian Hamiltonian $H_{eff}$ which neglects the corresponding jump terms, e.g., $H_{eff}=H-i\sum_{\mu}\hat{\Gamma}_{\mu}\hat{\Gamma}_{\mu}^{\dag}/2$~\cite{Ju}.

In order to solve Eq.(1), we rewrite the $N\times N$ density matrix $\rho(t)$ into a column vector, e.g., $|\hat{\rho}(t)\rangle=[\rho(1,:)^T,\rho(2,:)^T,\cdots,\rho(n,:)^T]$ by using the Hilbert-Schmidt expression of the density matrix~\cite{Havel,Am-Shallem,Scopa}. After a straightforward algebra, we can recast the non-Hermitian
Hamiltonian Lindblad equation given by Eq.(1) into a vectorized form
\begin{eqnarray}
\label{vectorized equation}
i\frac{\partial|\hat{\rho}(t)\rangle}{\partial t}&=&(\hat{\mathcal{H'}}+\hat{\mathcal{G'}})|\hat{\rho}(t)\rangle\\ \nonumber
&=&                                                                                                                                                                                                                                                                                                                                                                                                                                                                                                                                                                                                                                                                                                                                                                                                                                                                                                                                                                                                                                                                                                                                                                                                                                                                                                                                                                                                                                                                                                                                                                                                                                                                                                                                                                                                                                                                                                                                                                                                                                                                                                                                                                                                                                                                                                                                                                                                                                                                                                                                                                                                                                                                                                                                                                                                                                                                                                                                         \hat{L'}|\hat{\rho}(t)\rangle,
\end{eqnarray}
Obviously, the form of vectorized equation given by Eq.(3) is mathematically isomorphic to the time-dependent Schr$\ddot{o}$dinger equation. Denote $\hat{L'}=\hat{\mathcal{H'}}+\hat{\mathcal{G'}}$, here $\hat{\mathcal{H'}}=-i(\hat{H}\otimes I-I\otimes\hat{H}^\dag)$ and $\hat{\mathcal{G'}}=\sum_{n}\hat{\Gamma}_{\mu}\otimes\hat{\Gamma}_{\mu}^{\ast}-\frac{\hat{\Gamma}_{\mu}^{\dag}\hat{\Gamma}_{\mu}}{2}\otimes I-I\otimes\frac{(\hat{\Gamma}_{\mu}^{\dag}\hat{\Gamma}_{\mu})^{\ast}}{2}$ which are represented as $N^2\times N^2$ matrixes. However, it's worth mentioning that $\hat{L'}$ is including two different effects of the open system, one is derived from the non-Hermiticity of system with gain and loss, the other is arising from the dissipation of environment.

Within the vectorized framework,  the spectrum of the Liouvillian
superoperator $\hat{L'}$ describing the open system with decoherence and non-Hermiticity is easily obtained via the relation
\begin{equation}
\label{eigenvalues and eigenstates}
\mathcal{L'}|\hat{\rho}^{i}\rangle=\lambda_{i}|\hat{\rho}^{i}\rangle, (i=1,2,...n)
\end{equation}
where $\lambda_{i}$ and $|\hat{\rho}^{i}\rangle$ are the eigenvalues and eigenstates
of the Liouvillian $\mathcal{L'}$(a matrix form of the $\hat{L'}$), respectively. Generally speaking, for a time-independent Hermitian superoperator $\mathcal{L'}$, there is at least one steady state
if $\mathcal{L'}|\hat{\rho}^{s}\rangle=0$~\cite{Albert}. However, due to the non-Hermiticity of $\mathcal{L'}$, whose energy spectra are in general complex. For negative or positive imaginary part of the eigenvalues, the corresponding eigenstates will undergo exponential dissipation or amplification in time, while only the eigenstate with null eigenvalue survives~\cite{Mostafavi,Hatano}. Therefore, extending to the non-Hermitian case, there exist steady states if and only if at least one eigenvalues with zero-imaginary parts and the remanent ones with non-positive imaginary parts are fulfilled, equivalently,
\begin{subequations}
\begin{align}
Im[\lambda_{i}]=0,\quad%
\label{Mapl1}\\
\quad Im[\lambda_{j\neq i}]\leq 0.\label{Mapl2}%
\end{align}
\end{subequations}
The first equation ensures the existence of steady states, and the second one avoids the fact that the populations of the corresponding eigenstates grow exponentially. Comparing to the condition appeared~\cite{Albert} where the steady state $|\hat{\rho}^{s}\rangle\varpropto |\hat{\rho}^{1}\rangle$ is associated to $\lambda_{1}=0$ ( $|Re[\lambda_{1}]|<|Re[\lambda_{2}]|<...<|Re[\lambda_{n}]|$), our extension provides a more stringent constraint to find and classify a class of steady states in the open systems described by non-Hermitian Hamiltonian. In what following, we apply the vectorized Lindblad
equation to study an open two-level system with gain, loss and dissipation, and report on intriguing properties of non-Hermitian
systems including EPs, steady states and topological phases.
\begin{figure}[tbp] \centering
\includegraphics[width=4cm]{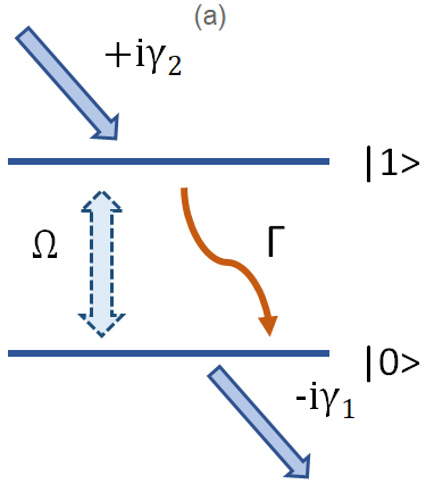}
\includegraphics[width=4cm]{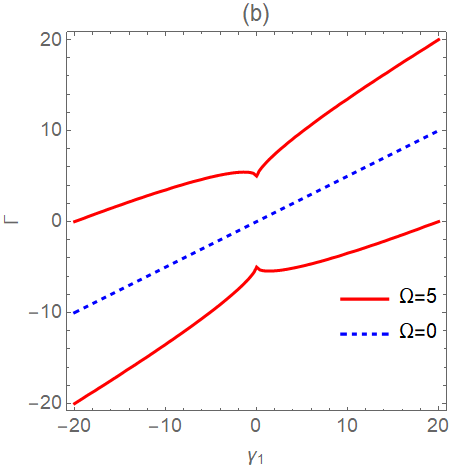}
 \caption{(Color online) (a) Scheme of an open two-level non-Hermitian system with gain $\gamma_{2}$ and loss $\gamma_{1}$ coupled to a dissipative environment. (b) Positions of exceptional points located in the parameter space for different $\Omega$. Other parameter is set $\gamma_{2}=\gamma_{1}$.
}\label{fig1}
\end{figure}

\section{An open two-level system with non-Hermiticity and dissipation}

For simplicity, we consider a two-level system consisting of lower level $|0\rangle$ and higher level $|1\rangle$ coupled to an environment with loss $\gamma_{1}$ , gain $\gamma_{2}$ and the dissipation rate $\Gamma$ decays from $|1\rangle$ to $|0\rangle$, $\Omega$ is the coupling strength
between two levels as shown Fig.1(a), which is the simplest and can be implemented in a variety of ways, including coupled quantum harmonic oscillators~\cite{Roccati}, waveguides~\cite{El-Ganainy} and microcavities~\cite{Peng0}. The Hamiltonian of system is described as
\begin{equation}
\label{H2}
\hat{H}=\frac{1}{2}\left(\begin{matrix}-i\gamma_{1}&\Omega\\\Omega&i\gamma_{2}\end{matrix}\right),
\end{equation}
combining with an amplitude damping term
\begin{eqnarray}
\label{decay}
\hat{\Gamma}_-=\Gamma\cdot\hat{\sigma}_{-}=\left(\begin{matrix}0&\Gamma\\0&0\end{matrix}\right).
\end{eqnarray}
Substituting Eqs.(6) and (7) into Eq.(3), we can obtain the Liouvillian superoperator $\hat{L'}$ with matrix form
 \begin{equation}
\label{vectorized form}
\mathcal{L'}=i\left(\begin{matrix}-\gamma_{1} & \frac{i\Omega}{2} & -\frac{i\Omega}{2} & \Gamma\\ \frac{i\Omega}{2} & \frac{\gamma_{2}-\gamma_{1}-\Gamma}{2}& 0&-\frac{i\Omega}{2}\\ -\frac{i\Omega}{2} & 0 & \frac{\gamma_{2}-\gamma_{1}-\Gamma}{2}&\frac{i\Omega}{2}\\0&-\frac{i\Omega}{2}&\frac{i\Omega}{2}&\gamma_{2}-\Gamma\end{matrix}\right),
\end{equation}
whose eigenvalues are
\begin{equation}
\begin{aligned}
\label{eigenvalues}
\lambda_{1}&=i\frac{1}{2}\eta_-,  \\
\lambda_{2}&=i\frac{1}{2}[\eta_{-}-2i\Lambda(0)], \\
\lambda_{3}&=i\frac{1}{2}[\eta_{-}+i\Lambda(\sqrt{3})],\\
\lambda_{4}&=i\frac{1}{2}[\eta_{-}+i\Lambda(-\sqrt{3})],
\end{aligned}
\end{equation}
and the corresponding eigenstates are given
\begin{equation}
\begin{aligned}
\label{eigenstates}
|\hat{\rho}^{1}\rangle&\propto\left(\begin{matrix} 0 \\ 1 \\ 1 \\ 0\end{matrix}\right),\\
|\hat{\rho}^{2}\rangle&\propto\left(\begin{matrix} 1-\frac{[\eta_{+}+2i\Lambda(0)](3\eta_{+}+\Theta^2-12 \Omega^2)}{6 \Omega^2 \Theta} \\ -\frac{i[\eta_{+}+2i\Lambda(0)]}{2 \Omega} \\ \frac{i[\eta_{+}+2i\Lambda(0)]}{2 \Omega} \\ 1\end{matrix}\right), \\
|\hat{\rho}^{3}\rangle&\propto\left(\begin{matrix} 1+\frac{[\eta_{+}-i\Lambda(\sqrt{3})][-i\Lambda(\sqrt{3})]}{2 \Omega^2} \\ -\frac{i[\eta_{+}-i\Lambda(\sqrt{ 3})]}{2 \Omega} \\ \frac{i[\eta_{+}-i\Lambda(\sqrt{3})]}{2 \Omega} \\ 1\end{matrix}\right),\\
|\hat{\rho}^{4}\rangle&\propto\left(\begin{matrix} 1+\frac{[\eta_{+}-i\Lambda(-\sqrt{3})][-i\Lambda(-\sqrt{3})]}{2 \Omega^2} \\ -\frac{i[\eta_{+}-i\Lambda(-\sqrt{ 3})]}{2 \Omega} \\ \frac{i[\eta_{+}-i\Lambda(-\sqrt{3})]}{2 \Omega} \\ 1\end{matrix}\right).
\end{aligned}
\end{equation}
Here we define $\Lambda(x)=\frac{(i+x)\Theta}{6}+\frac{(i-x)(\eta_+^2-4\Omega^2)}{2\Theta}$, and $\Theta=\left[54 \Gamma \Omega^2+\sqrt{2916\Gamma^2 \Omega^4-27(\eta_+^2-4\Omega^2)^3}\right]^{1/3}$, with
$\eta_{\pm}=\gamma_2\pm \gamma_1-\Gamma$. Since $\mathcal{L'}$ is not Hermitian, its eigenvalues are complex and the eigenvectors are, in general,
neither orthogonal nor normalized.

For $\Omega = 0$, there is an exceptional point located at $\gamma_{1}+\gamma_{2}=\Gamma$ where all the eigenvalues $\lambda_{i}$ are degenerate $-2i\gamma_{1}$ and the corresponding eigenstates $|\hat{\rho}^{i}\rangle$ coalesce to $(1,0,0,0)^T$. In contrast,
for $\Omega \neq 0$, the eigenvalues $\lambda_{3,4}$ are degenerated if only if $\Lambda(\sqrt{3})=\Lambda(-\sqrt{3})$, namely,
 \begin{equation}
\label{condition of EPs}
\eta_+=\pm\sqrt{3(4\Gamma^2\Omega^4)^{1/3}+4\Omega^2}.
\end{equation}
In this situation, the coalescence of the eigenvalues, along with its corresponding eigenmatrices $\rho_{3,4}$ respectively coalesce to

\begin{subequations}
\label{NLEPs}
\begin{equation} \lambda^{NLEPs}=i\frac{1}{2}[\eta_{-}-(2\Gamma\Omega^2)^{1/3}],\end{equation}
\begin{equation} |\hat{\rho}^{NLEPs}\rangle \propto\left(\begin{matrix} 1+\frac{[\eta_{+}+(2\Gamma\Omega^2)^{1/3}](2\Gamma\Omega^2)^{1/3}}{2 \Omega^2} \\ -\frac{i[\eta_{+}+(2\Gamma\Omega^2)^{1/3}]}{2 \Omega} \\ \frac{i[\eta_{+}+(2\Gamma\Omega^2)^{1/3}]}{2 \Omega} \\ 1\end{matrix}\right).\end{equation}
\end{subequations}
Unlike Lindblad exceptional points (LEPs) proposed in Refs.~\cite{Caspel,Arkhipov,Minganti3} where LEPs are only induced by the non-Hermiticity of the effective Hamiltonian, our degenerate points are jointly induced by the effects of non-Hermiticity and decoherence as shown in Fig.1(b), where the
exceptional lines represent the positions of EPs in the parameter space for different $\Omega$.  Interestingly, there exist double-exceptional points in the presented system for $\Omega\neq0$, in contrast to the case of $\Omega=0$. We call these EPs as non-Hermitian Lindblad exceptional points (NLEPs).

To further explore the properties of Liouvillian spectrum, according to Eq.(\ref{eigenvalues and eigenstates}), the second and third rows of the eigen-equation governed by Eq.(\ref{vectorized form}) read as
\begin{subequations}
\label{eigen_L_2,3}
\begin{equation}-\Omega\rho_{00}^{j}/2+i(\gamma_{2}-\gamma_{1}-\Gamma)\rho_{01}^{j}/2+\Omega\rho_{11}^{j}/2=\lambda_{j}\rho_{01}^{j}\end{equation}
\begin{equation}\Omega\rho_{00}^{j}/2+i(\gamma_{2}-\gamma_{1}-\Gamma)\rho_{10}^{j}/2-\Omega\rho_{11}^{j}/2=\lambda_{j}\rho_{10}^{j},\end{equation}
\end{subequations}
where $\rho_{mn}^{j}$ ($m,n=0,1$) are the elements of eigenstates $|\hat{\rho}^{j}\rangle$.
More specifically, the diagonal elements $\rho^{j}_{00,11}$ donate the probabilities of $|0\rangle$ and $|1\rangle$, respectively, while the off-diagonal elements $\rho^{j}_{01,10}$ represent the coherence between $|0\rangle$ and $|1\rangle$. If we restrict ourselves to the purely imaginary spectrum case, then Eq.(\ref{eigen_L_2,3}) reduces to
 \begin{equation}
\label{the classification of rho}
\rho_{01,10}^{j}=\mp\frac{i\Omega(\rho_{00}^{j}-\rho_{11}^{j})}{2\lambda_j-\eta_-}.
\end{equation}

The above formula indicates that the coherence elements $\rho_{01,10}^{j}$ can either be zero or purely imaginary numbers. If $\rho_{00}^{j}=\rho_{11}^{j}$ or $\Omega=0$ is fulfilled, and then $\rho_{01,10}^{j}$ are zero, which implying that there is incoherence between $|0\rangle$ and $|1\rangle$. Otherwise, there is coherence, and the phase between $|0\rangle$ and $|1\rangle$ must be $|\pi/2|$ owing to the fact that the off-diagonal elements $\rho_{01,10}^{j}$ are complex conjugate with each other for $Re[\lambda_{j}]=0$. In the following, we focus on the properties of steady states with and without coherence.

\subsection{Steady states without coherence}

\begin{figure}[tbp] \centering
\includegraphics[width=8.5cm]{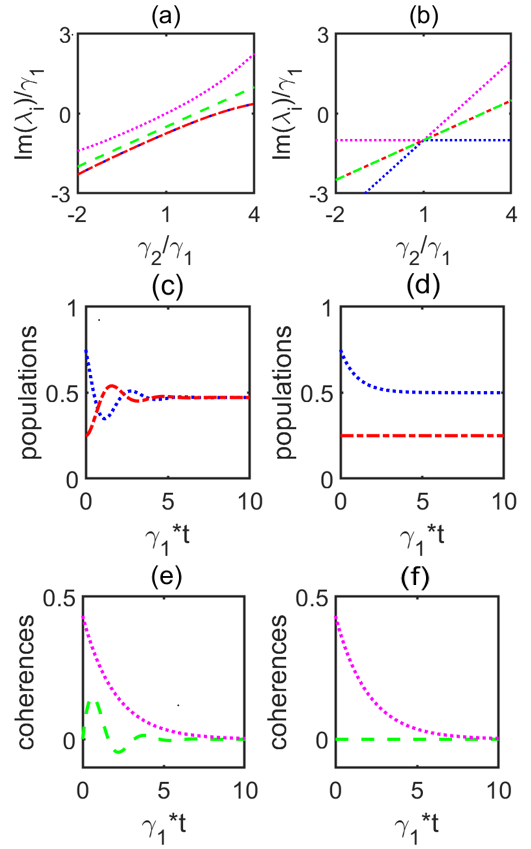}
\caption{(Color online) (a)(b)The imaginary part of the eigenvalues $\lambda_{1}$ (magenta dashed line), $\lambda_{2}$ (green dashed line), $\lambda_{3}$ (red dashed line) and $\lambda_{4}$ (blue dashed line) as a function of $\gamma_{2}/\gamma_{1}$. (c)(d) The probabilities of $\rho_{00}$(red dash-dot line), $\rho_{11}$ (blue dotted line) and (e)(f) the  coherences of $\rho_{01}$(magenta dashed line), $\rho_{10}$ (green dashed line) as the function of $\gamma_{1}t$  for the initial state $(1/4,\sqrt{3}/4,\sqrt{3}/4,3/4)^T$. Other parameters are chosen as (a)(c)(e) $\Omega=2\gamma_{1}$ and $\Gamma=\gamma_{1}$, (b)(d)(f) $\Omega=0$ and $\Gamma=2\gamma_{1}$. Support $\gamma_{1}$ is the energy unit in this work.}\label{fig2}
\end{figure}
First of all, let us consider there is incoherence between $|0\rangle$ and $|1\rangle$ when $\rho_{01,10}^{j}=0$, the eigen-equation governed by Eq.(\ref{vectorized form}) becomes
 \begin{equation}
\label{eigen_L_rho_2,3=0}
\left(\begin{matrix}i(\Gamma\rho_{11}^{j}-\gamma_{1}\rho_{00}^{j}) \\
\frac{\Omega}{2}(\rho_{11}^{j}-\rho_{00}^{j})  \\
\frac{\Omega}{2}(\rho_{00}^{j}-\rho_{11}^{j})  \\
i(\gamma_{2}-\Gamma)\rho_{11}^{j}\end{matrix}\right)=\lambda_j \left( \begin{matrix}\rho_{00}^{j} \\0\\0\\ \rho_{11}^{j}\end{matrix}
 \right).
\end{equation}
Suppose that $\rho_{00}^{j}=\rho_{11}^{j}$, we can conclude $\Gamma=\frac{1}{2}(\gamma_1+\gamma_2)$. In this case, there does exist a steady state at $\gamma_{2}=\gamma_{1}$ where the eigenvalue and the corresponding steady state are $\lambda_{s1}=0$ and $|\rho^{s1}\rangle=(\frac{1}{2},0,0,\frac{1}{2})^T$, respectively. Obviously, the state decays toward the steady state which is independent of the parameters of the open system. To further visualize the steady-state features, we plot the imaginary part of the eigenvalues $\lambda_{j}$ (the real part of the eigenvalues $Re[\lambda_{j}]=0$) as a function of $\gamma_{2}/\gamma_{1}$ for $\Gamma=\gamma_1$ in Fig.~\ref{fig2}(a). It can be clearly seen that the eigenvalue $\lambda_{1}$ becomes zero in the case of $\gamma_{2}/\gamma_{1}=1$, while the other eigenvalues in this condition are non-positive. Fig.~\ref{fig2}(c) shows the time evolution of the eigenstate as the function of $\gamma_{1}t$  when the initial state is prepared in $(1/4,\sqrt{3}/4,\sqrt{3}/4,3/4)^T$. As expected, the probabilities of $|0\rangle$ and $|1\rangle$ display a slow overall decay towards a well-defined value. On the other hand, when $\Omega=0$ is fixed, the elements of eigenstates satisfy the relation $\rho_{00}^{j}/\rho_{11}^{j}=\Gamma/(\gamma_{1}+\gamma_{2}-\Gamma)$ which implies the probabilities between $|0\rangle$ and $|1\rangle$ can be manipulated by continuously tuning parameters $\gamma_{1}$, $\gamma_{2}$ and $\Gamma$. In this case, there also exists a steady state at $\Gamma=\gamma_{2}$ where the eigenvalue and the corresponding steady state respectively become $\lambda_{s2}=i(\gamma_{2}-\Gamma)=0$ (the other eigenvalues at $\Gamma=\gamma_{2}$ are non-positive as shown Fig.~\ref{fig2}(b)) and $|\rho^{s2}\rangle=(\frac{\Gamma}{\gamma_1+\gamma_2},0,0,\frac{\gamma_1+\gamma_2-\Gamma}{\gamma_1+\gamma_2})^{T}$. Interestingly, this steady state is strongly depended on the parameters of the system, so that one can obtain the desired steady state by designing the parameters of the system as shown Fig.~\ref{fig2}(d). This property is completely different from the conventional dissipative systems where the steady states are uncontrollable~\cite{Albert,Nigro}.

It is worth mentioning that the aforementioned steady states are incoherence. The steady coherences rapidly decrease to zero, as shown Figs.2(e) and (f). Following, we will turn to the case of $\Omega \neq 0$, and demonstrate how the non-Hermiticity and environmental dissipative affect on the properties of steady states and topological phase.

\begin{figure*}[tbp] \centering
\includegraphics[width=17.5cm]{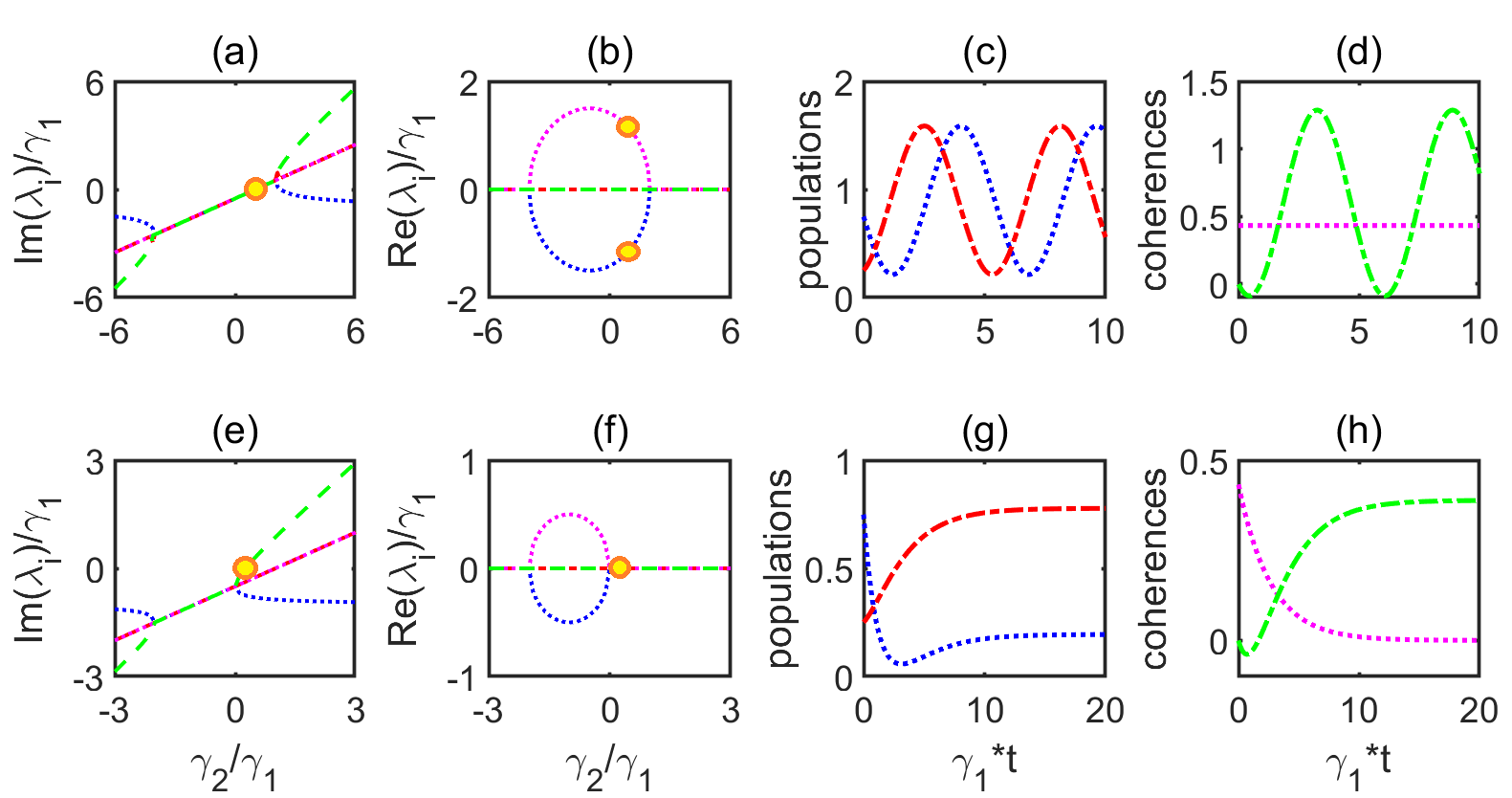}
 \caption{(Color online) The imaginary part and the real part of the eigenvalues $\lambda_{1}$ (magenta dashed line), $\lambda_{2}$ (green dashed line), $\lambda_{3}$ (red dashed line) and $\lambda_{4}$ (blue dashed line) as a function of $\gamma_{2}/\gamma_{1}$ for $\Gamma=0$ under different $\Omega$, e.g., $\Omega=1.5\gamma_{1}$ for (a)(b) and $\Omega=0.5\gamma_{1}$ for (e)(f).  The probabilities of $|0\rangle$ (blue line), $|1\rangle$ (red line) and the coherence of the real part (magenta line) and imaginary part (green line) as the function of $\gamma_{1}t$ when the initial state is prepared in $(1/4,\sqrt{3}/4,\sqrt{3}/4,3/4)^T$ under (c)(d) the PT-symmetric phase ($\Omega=1.5\gamma_{1}$) with $\gamma_{2}=\gamma_{1}$(Solid circles)and (g)(h) PT-symmetry broken phase ($\Omega=0.5\gamma_{1}$) with $\gamma_{2}=0.25\gamma_{1}$ ( Solid circles).}\label{fig3}
 \end{figure*}

\subsection{Steady states with coherence}

As we previously discussed, there exists coherence if only if $\rho_{00}\neq\rho_{11}$ and $\Omega\neq 0$. By ignoring the dissipative effect, namely, $\Gamma=0$, equation (8) reduces to a normal non-Hermitian system with gain and loss (whose eigenvalues and corresponding eigenstates are determined from Eqs.(9) and (10) by replacing $\Gamma=0$, respectively). In this case, the presented system exhibits PT-symmetry as $\gamma_{2}=\gamma_{1}$,  and the PT-symmetric phase
corresponds to $|\Omega|>\frac{\gamma_{2}+\gamma_{1}}{2}$, while $|\Omega|<\frac{\gamma_{2}+\gamma_{1}}{2}$,  the system is in the PT-symmetry broken phase. However, for $|\Omega|\gg\frac{\gamma_{2}+\gamma_{1}}{2}$, the system becomes a general closed hermitian system where only the real eigenvalues exist.

By exploiting Eq.(9), there are two exceptional points located at $\gamma_{2}/\gamma_{1}=\pm\frac{2\Omega}{\gamma_{1}}-1$. In particular, for $\gamma_{2}/\gamma_{1}\geq\frac{2\Omega}{\gamma_{1}}-1$ and $\gamma_{2}/\gamma_{1}\leq-\frac{2\Omega}{\gamma_{1}}-1$ cases, in which only the imaginary part of the eigenvalues survives while the real part is zero. This case is referred to as Fermi arcs~\cite{Bergholtz}. According to Eq.(14), if the eigenstates are located in Fermi arcs regime, the phase of eigenstates must be $|\pi/2|$.

To clearly support our theoretical predictions, we numerically depict the imaginary (real) part of the eigenvalues as a function of $\gamma_{2}/\gamma_{1}$ for $\Gamma=0$ under different $\Omega$, e.g., as shown in Figs.\ref{fig3}(a),(b) with $\Omega=1.5\gamma_{1}$ and Figs.\ref{fig3}(e),(f) with $\Omega=0.5\gamma_{1}$. Clearly, two exceptional points take place at $\gamma_{2}/\gamma_{1}=2$ and $\gamma_{2}/\gamma_{1}=-4$ (or $\gamma_{2}/\gamma_{1}=0$ and $\gamma_{2}/\gamma_{1}=-2$), respectively. For $\gamma_{2}/\gamma_{1}\geq2$ and $\gamma_{2}/\gamma_{1}\leq-4$ (or $\gamma_{2}/\gamma_{1}\geq0$ and $\gamma_{2}/\gamma_{1}\leq-2$), which are corresponding to Fermi arcs with $Re[\lambda_{i}]=0$ and $Im[\lambda_{i}]\neq0$.

However, it's worth mentioning that the exceptional points denoted by $\gamma_{2}/\gamma_{1}=\pm\frac{2\Omega}{\gamma_{1}}-1$ are located at Fermi arcs in the presented system. That implies the exceptional points are as a special case of Fermi arcs. In fact, the eigenstates near exceptional points with a topological phase $|\pi/2|$ had been verified in Refs.~\cite{Weidemann,Heiss3} but fail to demonstrate in Fermi arcs case. In our system, we show the eigenstates in Fermi arcs also have the topological phase $|\pi/2|$.

Interestingly, in the absence of $\Gamma$, there exists a steady state with coherence which is related to $2|\Omega|/(\gamma_{2}+\gamma_{1})$. When $|\Omega|>\frac{\gamma_{2}+\gamma_{1}}{2}$, there exists a steady state with PT-symmetry located at $\gamma_{2}=\gamma_{1}$, while $|\Omega|<\frac{\gamma_{2}+\gamma_{1}}{2}$, there exists a steady state without PT-symmetry located at $\gamma_{2}=0.25\gamma_{1}$.
Figs.\ref{fig3}(c),(d) demonstrate the probabilities and the coherences as the function of $\gamma_{1}t$ for $\Omega=1.5\gamma_{1}$ and $\gamma_{2}=\gamma_{1}$ when the initial state is $(1/4,\sqrt{3}/4,\sqrt{3}/4,3/4)^T$. As expected, the system tends to a dynamic equilibrium steady state with oscillating probabilities and coherences due to the fact that the system is in the PT-symmetric phase. By contrast, Figs.\ref{fig3}(g),(h) illustrate the time evolution of diagonal elements $\rho_{00}$ ($\rho_{11}$) donating probabilities of $|0\rangle$, $|1\rangle$ and the off-diagonal element $\rho_{01}$($\rho_{10}$) with the real part and imaginary part as the function of $\gamma_{1}t$ for $\Omega=0.5\gamma_{1}$ and $\gamma_{2}=0.25\gamma_{1}$ when the initial state is $(1/4,\sqrt{3}/4,\sqrt{3}/4,3/4)^T$. Clearly, the probabilities and the coherences tend to a stable value and the system ultimately reaches to a steady state with coherence.

\begin{figure}[tbp] \centering
\includegraphics[width=8.5cm]{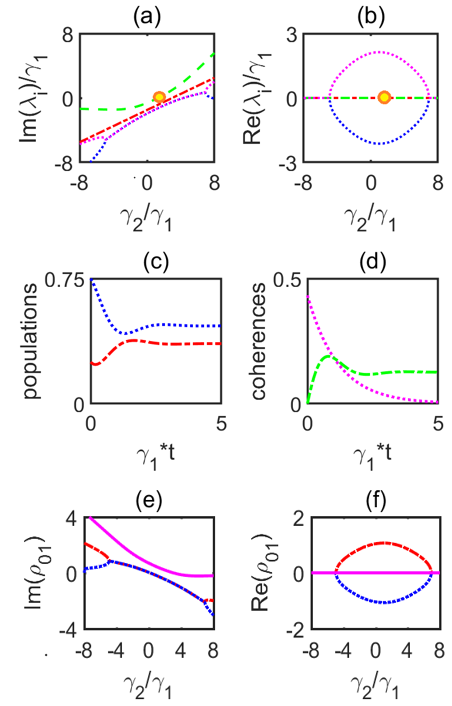}
 \caption{(Color online)(a) The imaginary part and (b) the real part of the eigenvalues $\lambda_{1}$ (magenta dashed line), $\lambda_{2}$ (green dashed line), $\lambda_{3}$ (red dashed line) and $\lambda_{4}$ (blue dashed line) as a function of $\gamma_{2}/\gamma_{1}$.  (c) The time evolution of diagonal elements $\rho_{00}$ ($\rho_{11}$) donating probabilities of $|0\rangle$ (blue line), $|1\rangle$ (red line) and (d) the off-diagonal element $\rho_{01}$($\rho_{10}$) with the real part (magenta line) and imaginary part (green line) as the function of $\gamma_{1}t$ when the initial state is $(1/4,\sqrt{3}/4,\sqrt{3}/4,3/4)^T$ for $\gamma_{2}=1.298\gamma_{1}$. (e) The imaginary part and (f) the real part of the $\rho_{01}^{2}$ (magenta solid line), $\rho_{01}^{3}$  (red dash-dot line) and $\rho_{01}^{4}$  (blue dotted line) as a function of $\gamma_{2}/\gamma_{1}$. Other parameters are chosen as $\Omega=2\gamma_{1}$ and $\Gamma=2\gamma_1$.}\label{fig4}
\end{figure}

On the other hand, taking the dissipative effect ($\Gamma\neq0$) into account, several distinct features can be found in Fig.4. Firstly, $\Gamma\neq0$ leads to a splitting of the eigenvalues at $\gamma_{2}/\gamma_{1}=\frac{\Gamma\pm \bigtriangleup}{\gamma_{1}}-1$, with $\bigtriangleup =\sqrt{3(4\Gamma^2\Omega^4)^{1/3}+4\Omega^2}$, as shown in Figs.\ref{fig4}(a) and \ref{fig4}(b). This result indicates that any environmental perturbation leads to a splitting of the eigenvalues that scales with the square root $\bigtriangleup$. Secondly, there exists a steady state with coherence. In order to illustrate this property of steady-state coherence clearly, the time evolution of populations $|0\rangle$ ($|1\rangle$) and coherence for initial eigenstate prepared in $(1/4,\sqrt{3}/4,\sqrt{3}/4,3/4)^T$ are depicted in Figs.\ref{fig4}(c) and \ref{fig4}(d), respectively. It is clearly seen, the diagonal elements $\rho_{00}$ and $\rho_{11}$ donating probabilities of $|0\rangle$ and $|1\rangle$  will eventually reach at a stable value, while the off-diagonal element $\rho_{01}$ representing coherence disappears gradually but the imaginary part of coherence steady state survives. Thirdly, there also exist Fermi arcs accompanied with $Re[\lambda_j]=0$ and $Im[\lambda_j]\neq0$ for $\gamma_{2}/\gamma_{1}\geq\frac{\Gamma+ \bigtriangleup}{\gamma_{1}}-1$ and $\gamma_{2}/\gamma_{1}\leq\frac{\Gamma- \bigtriangleup}{\gamma_{1}}-1$. At the same time, Fig.\ref{fig4}(e) and \ref{fig4}(f) show the off-diagonal elements $\rho_{01(10)}^{j}$ with $j=2,3,4$ (the eigenstate $\hat{\rho}^{1}$ is trivial) as a function of $\gamma_{2}/\gamma_{1}$ for $\Omega=2\gamma_{1}$ and $\Gamma=2\gamma_{1}$. As predicted by theory, $\rho_{01(10)}^{j}$ are purely imaginary numbers for $\gamma_{2}/\gamma_{1}\geq 6.92$ and $\gamma_{2}/\gamma_{1}\leq-4.92$.

Surprisingly, even though these eigenstates in Fermi arcs regime are suffered from the environmental dissipative effect, their off-diagonal elements always survive and the phases of eigenstates are still $|\pi/2|$. This result indicates these topological properties are robust against the dissipative environment. To better support our conclusion,
we resort to the topological properties of EPs which are as a special case of Fermi arcs. According to Eq.(12), the off-diagonal elements of $\hat{\rho}^{NLEPs}$ are always nonzero for $\gamma_{2}/\gamma_{1}\geq\frac{\Gamma+ \bigtriangleup}{\gamma_{1}}-1$ and $\gamma_{2}/\gamma_{1}\leq\frac{\Gamma- \bigtriangleup}{\gamma_{1}}-1$ regardless of the parameter $\Gamma$. This means the phase of $\hat{\rho}^{NLEPs}$ is immune to the effect of environmental decoherence and topologically protected.

\section{Conclusion}

In summary, we have explored an open two-level system with non-Hermitian and dissipation via mapping the non-Hermitian Lindblad equation to a vectorized equation by which the analytical expressions of the Liouvillian spectrum are analytically obtained. In particular, we have proposed a general condition of the steady state in the open non-Hermitian systems. Our results show that the combination of decoherence and non-Hermiticity
effects lead to unexpected behavior that differs significantly from the usual dynamical picture. Specifically, the properties of steady states and EPs are determined by  the interplay between non-Hermiticity and decoherence. Additionally, the relationship between Fermi arcs and EPs has been unveiled. We find that EPs are as a special case of Fermi arcs at which the eigenstates located have a topological phase $|\pi/2|$ which is not affected by the dissipative environment. Therefore, this present study is expected to deepen the understanding of the open systems with decoherence and non-Hermiticity effects.

\acknowledgements
This work is supported by the National Natural Science Foundation of China (Grants No.11747107), the Natural Science Foundation of Guangdong province (Grants No. 2017A030307023), the Scientific Research Project of Hunan Province Department of Education (Grant Nos.19B060). Y N Guo is supported by the Program of Changsha Excellent Young Talents (kq2009076 and kq2106029).


\begin{thebibliography}{99}

\bibitem{Breuer}H. Breuer, and F. Petruccione, The Theory of Open Quantum Systems (Oxford University Press, Oxford, 2007).

\bibitem{Kwiat}P. G. Kwiat, S. Barraza-Lopez, A. Stefanov, and N. Gisin, Experimental entanglement distillation and hidden non-locality. Nature \textbf{409},1014 (2001).

\bibitem{Lidar}D. A. Lidar, I. Chuang, and K. B. Whaley, Decoherence free subspaces for quantum computation. Phys. Rev. Lett. \textbf{81}, 2594(1998).

\bibitem{Facchi}P. Facchi, D. A. Lidar, and S. Pascazio, Unification of dynamical decoupling and the quantum Zeno effect. Phys. Rev. A \textbf{69}, 032314 (2004).

\bibitem{Heiss}W. D. Heiss, and H.L. Harney, The chirality of exceptional points, Eur. Phys. J. D \textbf{17}, 149 (2001)



\bibitem{Doppler}M. Miri, and A. Al$\grave{u}$, Exceptional points in optics and photonics, Science \textbf{363}, 7709 (2019).

\bibitem{Dembowski}C. Dembowski, B. Dietz, H. D. Gr$\ddot{a}$f, H. L. Harney, A. Heine, W. D. Heiss, and A. Richter, Observation of a chiral state in a microwave cavity, Phys. Rev. Lett. \textbf{90}, 034101 (2003).
\bibitem{Chen}W. Chen, S. K. $\ddot{O}$zdemir, G. Zhao, J. Wiersig, and L. Yang, Exceptional points enhance sensing in an optical microcavity, Nature (London) \textbf{548}, 192 (2017).

\bibitem{Zhang}J. Zhang, B. Peng, S. K. $\ddot{O}$zdemir, K. Pichler, D. O. Krimer, G. Zhao, F. Nori, Y. X. Liu, S. Rotter, and L. Yang, A phonon laser operating at an exceptionalpoint, Nature Photonics \textbf{12}, 479 (2018).

\bibitem{Dora}B. D$\grave{o}$ra, M. Heyl, and R. Moessner, The Kibble-Zurek mechanism at exceptional points, Nat. Common. \textbf{10}, 2254 (2019).

\bibitem{Cerjan}A. Cerjan, S. Huang, K. P. Chen, Y. D. Chong, and M. C. Rechtsman, Experimental realization of a Weyl exceptional ring, Nature Photonics \textbf{13} 623 (2019)

\bibitem{Kawabata}S. K. $\ddot{O}$zdemir, S. Rotter, F. Nori, and L. Yang, Parity-time symmetry and exceptional points in photonics, Nature Materials \textbf{18}, 783 (2019).



\bibitem{Yoshida}T. Yoshida, R. Peters, N. Kawakami, and Y. Hatsugai, Symmetry-protected exceptional rings in two-dimensional correlated systems with chiral symmetry, Phys. Rev. B \textbf{99}, 121101(R) (2019).
 \bibitem{Weidemann}J. Doppler, A. A. Mailybaev, J. B$\ddot{o}$hm, U. Kuhl, A. Girschik, F. Libisch, T. J. Milburn, P. Rabl, N. Moiseyev, and S. Rotter, Dynamically encircling an exceptional point for asymmetric mode switching, Nature \textbf{537}, 76 (2016).


\bibitem{Lin}Y. C. Lee, M. H. Hsieh, S. T. Flammia, and R. K. Lee, Local PT-symmetry violates the no-signaling principle, Phys. Rev. Lett. \textbf{112}, 130404 (2014).
\bibitem{Regensburger}G. Japaridze, D. Pokhrel, and X. Q. Wang, No-signaling principle and Bell inequality in PT-symmetric quantum mechanics, J. Phys. A: Math. Theor.\textbf{50}, 18503 (2017).
\bibitem{Brandstetter}M. Brandstetter, M. Liertzer, C. Deutsch, P. Klang, J. Schoberl, H. E. Tureci, G. Strasser, K. Unterrainer,
and S. Rotter, Reversing the pump dependence of a laser at an exceptional point, Nat. Commun. \textbf{5}, 4034 (2014).
\bibitem{Peng} B. Peng, S. K. $\ddot{O}$zdemir, S. Rotter, H. Yilmaz, M. Liertzer, F. Monifi, C. M. Bender, F. Nori, and L. Yang, Loss-induced suppression and revival of lasing, Science \textbf{346}, 328 (2014).
\bibitem{Huang}C. M. Bender, D. C. Brody, H. F. Jones, and B. K. Meister, Faster than Hermitian quantum mechanics, Phys. Rev. Lett. \textbf{98}, 040403 (2007).





\bibitem{Hodaei}S. Dey, A. Fring, and T. Mathanaranjan, Non-Hermitian systems of Euclidean Lie algebraic type with real eigenvalue spectra, Ann. Phys.\textbf{346}, 28 (2014).

\bibitem{Xiao}K. Zelaya, S. C. Cruz, and O. Rosas-Ortiz, On the construction of non-Hermitian Hamiltonians with all-real spectra through supersymmetric algorithms, Geomet. Meth. Phys.\textbf{XXXVIII}, 283 (2020).

\bibitem{Chen2}P. Y. Chen, M. Sakhdari, M. Hajizadegan, Q.S. Cui, M. M. Cheng, R. E. Ganainy, and A. Al$\grave{u}$, Generalized parity time symmetry condition for enhanced sensor telemetry, Nat. Electron. \textbf{1} 297 (2018).

\bibitem{Ozdemir}K. Kawabata, K. Shiozaki, M. Ueda, and M. Sato, Symmetry and topology in non-Hermitian physics, Phys. Rev. X \textbf{9}, 041015 (2019).

\bibitem{Ghatak}J. da Providencia, N. Bebiano, and J. P. da Providencia, Non-Hermitian Hamiltonians with real spectrum in quantum mechanics, Braz. J. Phys.\textbf{41} 78 (2011).

\bibitem{Chu}Y. M. Chu, Y. Liu, H. Liu, and J. M. Cai, Quantum sensing with a single-qubit pseudo-Hermitian system, Phys. Rev. Lett. \textbf{124}, 020501 (2020).
\bibitem{AGuo}A. Guo, G. J. Salamo, D. Duchesne, R. Morandtti, M. Volatier-Ravat, V. Aimez, G. A. Siviloglou, and D. N. Christodoulides, Observation of PT-Symmetry breaking in complex optical potentials, Phys. Rev. Lett. \textbf{103}, 093902 (2009).



\bibitem{Bergholtz}E. J. Bergholtz, J. C. Budich, and F. K. Kunst, Exceptional topology of non-Hermitian Systems, Rev. Mod. Phys. \textbf{93}, 015005 (2021).
\bibitem{Berry}K. Kawabata, T. Bessho, and M. Sato, Non-Hermitian topology of exceptional points, arXiv:1902.08479




\bibitem{Heiss2}D. Heiss, Mathematical physics: Circling exceptionalpoints, Nat. Phys. \textbf{12}, 823 (2016).
\bibitem{Wang}H. L. Wang, L. J. Lang, and Y. D. Chong, Non-Hermitian dynamics of slowly varying Hamiltonians, Phys. Rev. A \textbf{98}, 012119 (2018).

\bibitem{Kozii}V.Kozii, L. Fu, Non-Hermitian topological theory of finite-lifetime quasiparticles: Prediction of bulk Fermi arc due to exceptional point, arXiv:1708.054841.
\bibitem{Zhou}H. Zhou, C.Peng, Y.Yoon, C. W. Hsu, K. A Nelson, L. Fu, J. D. Joannopoulos, M. Solja, B. Zhen, Observation of bulk Fermi arc and
polarization half charge from paired exceptional points, Science \textbf{359}, 1009 (2018).

\bibitem{Bessho}T. Bessho, K. Kawabata, and M. Sato, Topological classificaton of non-Hermitian gapless phases: Exceptional points and bulk Fermi arcs, JPS Conf. Proc. \textbf{30}, 011098 (2020).

\bibitem{Carlstrom}J. Carlstr$\ddot{o}$m, M. Stalhammar, J. C. Budich, E. J. Bergholtz, Knotted non-Hermitian metals, Phys. Rev. B 99, 161115 (2019).




\bibitem{Johan}J. Carlstr$\ddot{o}$m, Correlations in non-Hermitian systems and diagram techniques for the steady state, Phys. Rev. Res. \textbf{2}, 013078 (2020).
\bibitem{Pei}W. Pei, and X. L Gao, Connecting dynamical quantum phase transitions and topological steady-state transitions by tuning the energy gap, Phys. Rev. A \textbf{97}, 023627 (2018).

\bibitem{Mostafavi}F. Mostafavi, L. Q. Yuan, and H. Ramezani, Eigentates transition without undergoing an adiabatic process, Phys. Rev. Lett. \textbf{122}, 050404 (2019).


\bibitem{Caspel}F. Minganti, A. Miranowicz, R. W. Chhajlany, and F. Nori, Quantum exceptional points of non-Hermitian Hamiltonians and Liouvillians: The effects of quantum jumps, Phys. Rev. A \textbf{100}, 062131 (2019).

\bibitem{Arkhipov}I. I. Arkhipov, A. Miranowicz, F. Minganti, and F. Nori, Quantum and semiclassical exceptional points of a linear system of coupled cavities with losses and gain within the Scully-Lamb laser theory, Phys. Rev. A \textbf{101}, 013812 (2020).

\bibitem{Minganti3}F. Minganti, A. Miranowicz, R. W. Chhajlany, I. I. Arkhipov, and F. Nori, Hybrid-Liouvillian formalism connecting exceptional points of non-Hermitian Hamiltonians and Liouvillians via postselection of quantum trajectories, Phys. Rev. A \textbf{101}, 062112 (2020).



 \bibitem{Havel}T. F. Havel, Robust procedures for converting among Lindblad, Kraus and matrix representations of quantum dynamical semigroups, J. Math. Phys. \textbf{44}, 534 (2003).

\bibitem{Am-Shallem} M. Am-Shallem, A. Levy, I. Schaefer, and R. Kosloff, Three approaches for representing Lindblad dynamics by a matrix-vector notation,
 arXiv:1510.08634.


\bibitem{Scopa} S. Scopa, G. T. Landi, A. Hammoumi, and D. Karevski, Exact solution of time-dependent Lindblad equations with closed algebras, Phys. Rev. A, \textbf{99} 022105 (2019).

\bibitem{Hatano}N. Hatano, Exceptional points of the Lindblad operator of a two-level system, Mol.Phys. \textbf{117}, 2121 (2019).

 \bibitem{Walls}D. F. Walls, and G. J. Milburn, Quantum Optics (Springer, Berlin, 1994).

\bibitem{Ju}C. Y. Ju, A. Miranowicz, G. Y. Chen, and F. Nori, Quantum exceptional points of non-Hermitian Hamiltonians and Liouvillians: The effects of quantum jumps, Phys. Rev. A \textbf{100}, 062131 (2019).

\bibitem{Albert}V. V. Albert, and L. Jiang, Symmetries and conserved quantities in Lindblad master equations, Phys. Rev. A \textbf{89}, 022118 (2014).


 \bibitem{Roccati}F. Roccati, S. Lorenzo1, G M. Palma1, G. T Landi, M. Brunelli, and F Ciccarello, Quantum correlations in PT-symmetric systems, Quantum Sci. Technol. \textbf{6}, 025005 (2021).
 \bibitem{El-Ganainy}C. E. R$\ddot{u}$ter, K. G. Makris, R. El-Ganainy, D. N. Christodoulides, M. Segev, and D. Kip, Observation of parity-time symmetry in
optics, Nat. Phys. \textbf{6},192 (2010).
 \bibitem{Peng0}B. Peng, Parity-time-symmetric whispering-gallery microcavities, Nat. Phys. \textbf{10}, 394 (2014).


\bibitem{Nigro}D. Nigro, On the uniqueness of the steady-state solution of the Lindblad-Gorini-Kossakowski-Sudarshan equation, J. Stat. Mech. 043202 (2019).


\bibitem{Heiss3}W. D. Heiss, The physics of exceptional points, J. Phys. A: Math. Theor. \textbf{45}, 444016 (2012)





\end{thebibliography}
\end{document}